\documentclass[prd,nofootinbib,preprint,superscriptaddress,twocolumn,10pt]{revtex4}
\pdfoutput=1
\usepackage[utf8]{inputenc}
\usepackage{amsmath,amssymb}
\usepackage{epsfig}
\usepackage{comment}
\usepackage{bbm}
\usepackage{graphicx}
\usepackage[usenames,dvipsnames]{color}
\usepackage{bbold}
\usepackage[colorlinks,citecolor=blue]{hyperref}
\usepackage[normalem]{ulem}
\usepackage{multirow}

\usepackage{color}
\usepackage{subfigure}

\newcommand{\beq}{\begin{equation}}
\newcommand{\eeq}{\end{equation}}
\newcommand{\nn}{\nonumber}
\newcommand{\half}{\tfrac{1}{2}}
\newcommand{\vev}{{VEV}~}
\newcommand{\vevs}{{VEVs}~}
\usepackage{orcidlink}

\begin{document}
\title{Charged scalar portal dark matter with $U(1)_Y$ non-restoration and gravitational waves}


\author{Debasish Borah\orcidlink{0000-0001-8375-282X}}
\email{dborah@iitg.ac.in}
\affiliation{Department of Physics, Indian Institute of Technology Guwahati, Assam 781039, India}

\author{Dhruv Ringe\orcidlink{0000-0001-9052-418X}}
\email{ringedh@iitg.ac.in}
\affiliation{Department of Physics, Indian Institute of Technology Guwahati, Assam 781039, India}

\author{Indrajit Saha\orcidlink{0000-0002-7459-0838}}
\email{s.indrajit@iitg.ac.in}
\affiliation{Department of Physics, Indian Institute of Technology Guwahati, Assam 781039, India}

\begin{abstract}
We explore the possibility of reviving scalar singlet dark matter (DM) with a minimal extension by a charged singlet scalar which also leads to hypercharge non-restoration in the early Universe.  
The minimal scalar singlet DM annihilating via the Higgs portal is tightly constrained by direct-detection experiments, leaving a desert between two narrow regions of the DM mass near the Higgs resonance and around a TeV. The addition of a charged scalar portal revives this desert region while keeping the one-loop direct-detection rate within reach of current and future experiments. A non-zero vacuum expectation value of the charged scalar at high temperature can also drive $U(1)_Y$ hypercharge symmetry non-restoration in the early Universe. In a subset of the parameter space, the charge-breaking minimum gets restored to the standard Higgs minimum via a strong first-order phase transition, making it a viable scenario for the generation of baryon asymmetry of the Universe via leptogenesis. A large part of the parameter space consistent with DM relic, direct-detection, symmetry non-restoration and other experimental bounds can be probed at upcoming GW detectors such as LISA, BBO and DECIGO. 
\end{abstract}

\maketitle

\section{Introduction}
Numerous observations in astrophysics and cosmology suggest the presence of a non-luminous and non-baryonic form of matter, popularly known as dark matter (DM) contributing approximately five times of ordinary baryonic matter to the energy budget of the present Universe\cite{Planck:2018vyg, ParticleDataGroup:2024cfk, Cirelli:2024ssz}. As the standard model (SM) of particle physics does not explain the fundamental or particle origin of DM, several beyond standard model (BSM) proposals have been put forward to explain the origin of DM. Among them, the weakly interacting massive particle (WIMP) \cite{Kolb:1990vq, Jungman:1995df, Bertone:2004pz} has been the most popular one. In a generic WIMP framework, a particle DM candidate has sizeable non-gravitational interactions with the SM bath due to which it gets thermally produced in the early Universe with its relic set by thermal freeze-out. The same non-gravitational interactions between DM and the SM bath can also lead to observable DM-nucleon scattering at terrestrial detectors. However, null results at direct-detection experiments like LUX-ZEPLIN (LZ) ~\cite{LZ:2024zvo}, XENONnT \cite{XENON:2025vwd}, PandaX-4T \cite{PandaX:2024qfu} have already ruled out a large part of the parameter space of the simplest WIMP models like the scalar singlet DM \cite{Silveira:1985rk, McDonald:1993ex, Burgess:2000yq}.

The phenomenology of scalar singlet DM is primarily decided by two free parameters namely, the DM mass and its coupling with the SM Higgs. The above-mentioned constraints from direct-detection together with the constraints on Higgs invisible decay \cite{ATLAS:2023tkt, CMS:2023sdw} already rules out most of the relic density allowed parameter space leaving a tiny region of parameter space near the Higgs resonance and near a TeV \cite{GAMBIT:2017gge}. The intermediate desert region can be revived with the inclusion of additional scalars which can mediate DM annihilations without enhancing the direct-detection rates. Several earlier works considered additional neutral scalar singlets \cite{Casas:2017jjg} or higher electroweak multiplet scalars \cite{Bhattacharya:2016qsg, Campbell:2016zbp} to revive the scalar singlet DM. Here we extend the minimal scalar singlet DM model by a charged singlet scalar. This opens up new annihilation of DM into charged scalars reviving the previously disallowed region of DM masses. While the tree level DM-nucleon scattering can be kept sub-dominant by suitable choice of Higgs portal coupling, the charged scalar portal coupling of DM gives rise to DM-nucleon interactions at one-loop. Interestingly, a large part of the relic allowed parameter space of charged scalar portal singlet DM can saturate the latest bound on spin-independent DM-nucleon scattering rate, keeping the scenario verifiable in near future. 

The extension of the scalar singlet DM model by a charged scalar also allows the possibility of $U(1)_Y$ hypercharge symmetry non-restoration in the early Universe. This occurs due to a non-zero vacuum expectation value (VEV) acquired by the charged scalar at high temperatures which breaks the $U(1)_Y$ gauge symmetry. If the transition from the $U(1)_Y$ breaking minimum to the Electroweak (EW) minimum happens via a first-order phase transition (FOPT), this scenario can have potentially interesting consequences for the generation of the baryon asymmetry of the Universe \cite{Lopez-Zurdo:2023xmr} and primordial magnetic fields \cite{Grasso:2000wj}. In this paper, we focus on the possibility of achieving FOPT along with non-restoration, which brings additional detection prospects via observation of stochastic gravitational waves (GWs).
In the minimal setup of charged scalar portal DM, this FOPT occurs in a subset of the relic-allowed parameter space. As the charged singlet scalar plays non-trivial role in DM relic, DM-nucleon scattering as well as $U(1)_Y$ non-restoration and the related FOPT, we find interesting correlations between different observables. A large part of the parameter space can be probed at variety of future experiments searching for DM-nucleon scattering, stochastic GWs, charged lepton flavor violation and charge scalar decays at colliders.

This paper is organized as follows. In section \ref{sec:2}, we outline our model with the new particle spectrum, interactions and relevant theoretical constraints. In section \ref{sec:dm}, we discuss the details of dark matter followed by the details of symmetry non-restoration and FOPT in section \ref{sec:3}. In section \ref{sec:det} we briefly mention the detection prospects of the model and finally conclude in section \ref{sec:conc}.

\section{The Model}
\label{sec:2}
We consider a real singlet scalar DM $\eta$ which transforms non-trivially under an unbroken $Z_2$ symmetry while all other fields are $Z_2$-even. An additional $SU(2)_L$ singlet scalar $\phi$ with hypercharge $Y=+1$ (electromagnetic charge $q_{\rm{EM}}=+1$) is introduced to revive the singlet scalar DM parameter space. The corresponding scalar potential invariant under SM gauge and $Z_2$ symmetries can be written as
\begin{align}
    V(\eta, \phi, H) & = \frac{\mu^2_1}{2}\eta^2 + \mu^2_2 (\phi^\dagger \phi) -\mu^2_3 (H^\dagger H)+\frac{\lambda_1}{4} \eta^4 \nonumber \\
   & +\lambda_2 (\phi^\dagger \phi)^2 + \lambda_3 (H^\dagger H)^2 + \frac{\lambda_{12}}{2} \eta^2 (\phi^\dagger \phi) \nonumber \\
 &  + \frac{\lambda_{13}}{2} \eta^2 (H^\dagger H) + \lambda_{23} (\phi^\dagger \phi)( H^\dagger H)\,
 \label{eq:pot}
\end{align}
where $H$ is the SM Higgs doublet.
We write the Higgs doublet $H$ as 
\begin{align}
    H = \begin{bmatrix}
    G^+ \\
    \frac{1}{\sqrt{2}}(h+i G_0)
    \end{bmatrix}\,,
\end{align}
and the charged scalar $\phi$ as 
\beq
\phi = \frac{1}{\sqrt{2}}\,(\varphi+ i\xi)\,,
\eeq
where $G^{\pm},~G_0$ are the SM Goldstone bosons and $\varphi$ and $\xi$ are, respectively, the real scalar and pseudoscalar components of $\phi$. Demanding that the tree-level potential is bounded from below yields the following conditions
\beq
\begin{aligned}
&\lambda_i \ge 0 \quad (i=1,2,3)\,, \\
&\lambda_{ij} \ge -2\sqrt{\lambda_i\lambda_j}
\quad (ij \equiv 12,13,23)\,, \\
&\sqrt{\lambda_1\lambda_2\lambda_3}
+\frac{1}{2}\!\left(
\lambda_{12}\sqrt{\lambda_3}
+\lambda_{13}\sqrt{\lambda_2}
+\lambda_{23}\sqrt{\lambda_1}
\right)\! \ge 0 .
\end{aligned}
\eeq
Additionally, perturbativity requires $|\lambda|<4\pi$ for all quartic couplings of the scalar potential in Eq. \eqref{eq:pot}. 

A non-zero \vev of $\eta$ at the EW minimum would spontaneously break the $Z_2$ symmetry of the potential, leading to the possibility of $\eta$ decaying into SM particles. Hence $\eta$ \vev must vanish. The \vev of the charged scalar $\phi$ can be chosen to lie along the $\varphi$ direction via a $U(1)$ rotation. However, at the EW minimum, the $\varphi$ \vev must also vanish as charge-breaking minima are forbidden by $U(1)_{\rm{EM}}$ invariance. Hence, the low-temperature EW minimum of the potential lies along
$$(\langle \eta\rangle,\langle\varphi\rangle,\langle h\rangle) = (0,0,v)\,,$$ 
where $v=246\,$GeV. The existence of the EW minimum requires
\beq
\mu_3^2 =\lambda_3 v^2> 0\,,
\eeq
The potential could have additional local minima along $(\langle \eta\rangle,\langle\varphi\rangle,\langle h\rangle) = (v_{\eta},0,0),~ (0,v_{\varphi},0)$, or $(v_{\eta},v_{\varphi},0)$. In such cases, the EW minimum must be the global minimum, i.e.
\begin{align}
    V(0,0,v)&<V(v_\eta,0,0)\,,\\
    V(0,0,v)&<V(0, v_{\varphi},0)\,,\\
    V(0,0,v)&<V(v_\eta,v_{\varphi},0)\,.
\end{align}
These lead to the following tree-level constraints relating the parameters of the scalar potential
\begin{align}
\frac{\mu_3^4}{\lambda_3} &> \frac{\mu_1^4}{\lambda_1},~\text{if } \mu_1^2 < 0
~\text{(}\eta~\text{only minimum)}, \\[2mm]
\frac{\mu_3^4}{\lambda_3} &> \frac{\mu_2^4}{\lambda_2},
~\text{if } \mu_2^2 < 0
~ \text{(}\varphi~\text{only minimum)}, \\[2mm]
\frac{\mu_3^4}{\lambda_3} & >\frac{\lambda_2 \mu_1^4
+\lambda_1 \mu_2^4-\lambda_{12}\mu_1^2\mu_2^2}
{\lambda_1\lambda_2-\lambda_{12}^2/4},\nn\\
&\text{if }\lambda_1\lambda_2-\lambda_{12}^2/4>0 ~\text{(mixed minimum)}.
\end{align}

The tree-level masses of the physical scalars in the EW minimum of the potential are
\begin{align}
    m^2_{\eta} &= \mu_1^2+\frac{\lambda_{13}}{2}v^2\label{eq: meta}\,,\\
    m^2_{\phi^{\pm}} &= \mu_2^2+\frac{\lambda_{23}}{2}v^2\label{eq: mphi}\,,\\
    m^2_{h} &= -\mu_3^2+3\lambda_3 v^2 = 2\lambda_3 v^2\label{eq: mh}\,,
\end{align}
where $m_h = 125$\,GeV is the SM Higgs mass, yielding $\lambda_3 \approx 0.129$.
Eqs.\eqref{eq: meta}, \eqref{eq: mphi} and \eqref{eq: mh} can be used to trade the mass parameters for physical masses. 

\section{Dark matter}
\label{sec:dm}
The relic abundance of DM can be estimated by solving the Boltzmann equation
\begin{equation}
    \frac{dY_{\rm DM}}{dx}=-\frac{s(m_{\rm DM})}{x^2  \mathcal{H}(m_{\rm DM})}  \langle\sigma_{\rm ann} v\rangle (Y^2_{\rm DM}-(Y^{\rm eq}_{\rm DM})^2),
\end{equation}
where $Y_{\rm DM}=n_{\rm DM}/s$ is the comoving number density of DM with $n_{\rm DM}, s$ being DM number density and entropy density of the Universe respectively. $x=m_{\rm DM}/T \equiv m_\eta/T$ and $\mathcal{H}$ denotes the Hubble expansion parameter. $\langle\sigma_{\rm ann} v\rangle$ denotes the thermal averaged cross-section \cite{Gondolo:1990dk} for DM annihilation into lighter SM particles
\beq
\langle\sigma v\rangle = \frac{1}{8 m^4 T K_2^2\left(\frac{m}{T}\right)}\int_{4m^2}^{\infty}\sigma\,(s-4m^2)\sqrt{s}K_1\left(\frac{\sqrt{s}}{T}\right) ds \nonumber 
\eeq
where $K_i(x)$ is the modified Bessel function of $i$-th kind.

In addition to the usual Higgs portal annnihilation channels controlled by the quartic coupling $\lambda_{13}$, we also have $\eta\eta\rightarrow\phi\phi$ via the $\lambda_{12}$ or charged-scalar portal coupling. The cross-section is given by
\beq
\sigma = \frac{\lambda^2_{12}}{16\pi s}\sqrt{\frac{s-m_\phi^2}{s-m_\eta^2}}\,.
\label{eq:etaphi}
\eeq
We use \texttt{micrOMEGAs} \cite{Belanger:2026asz,Alguero:2023zol} to compute the relic density of dark matter by incorporating all the annihilation channels.

\begin{figure*}
    \centering
    \includegraphics[width=0.45\linewidth]{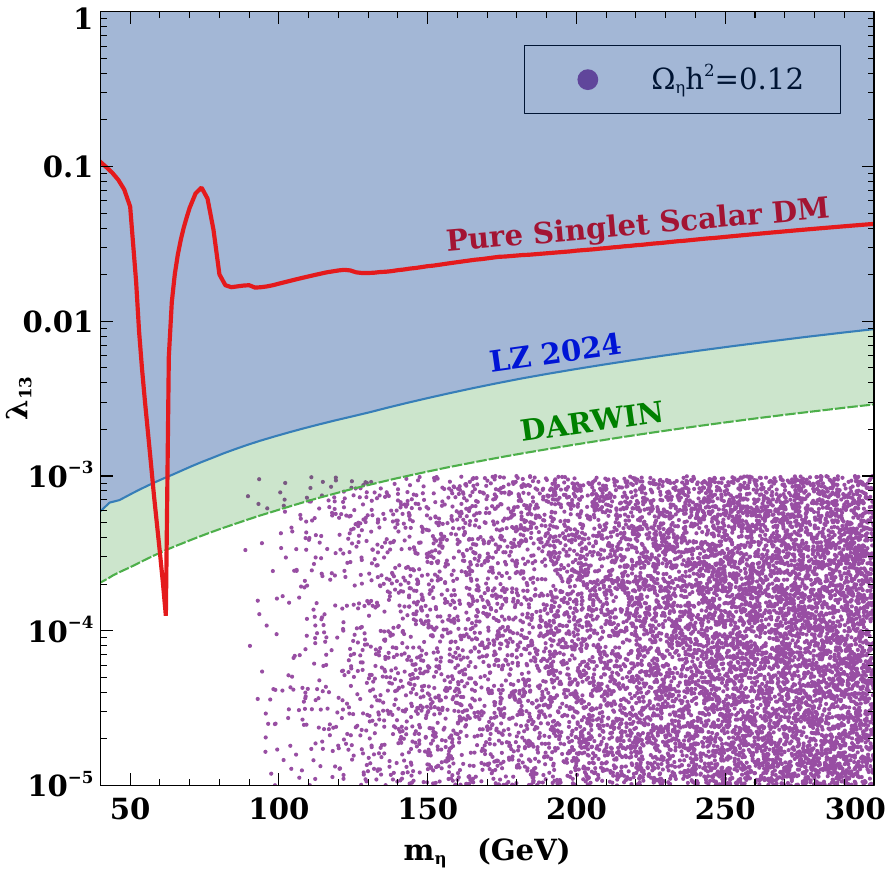}
    \hfill
     \includegraphics[width=0.45\linewidth]{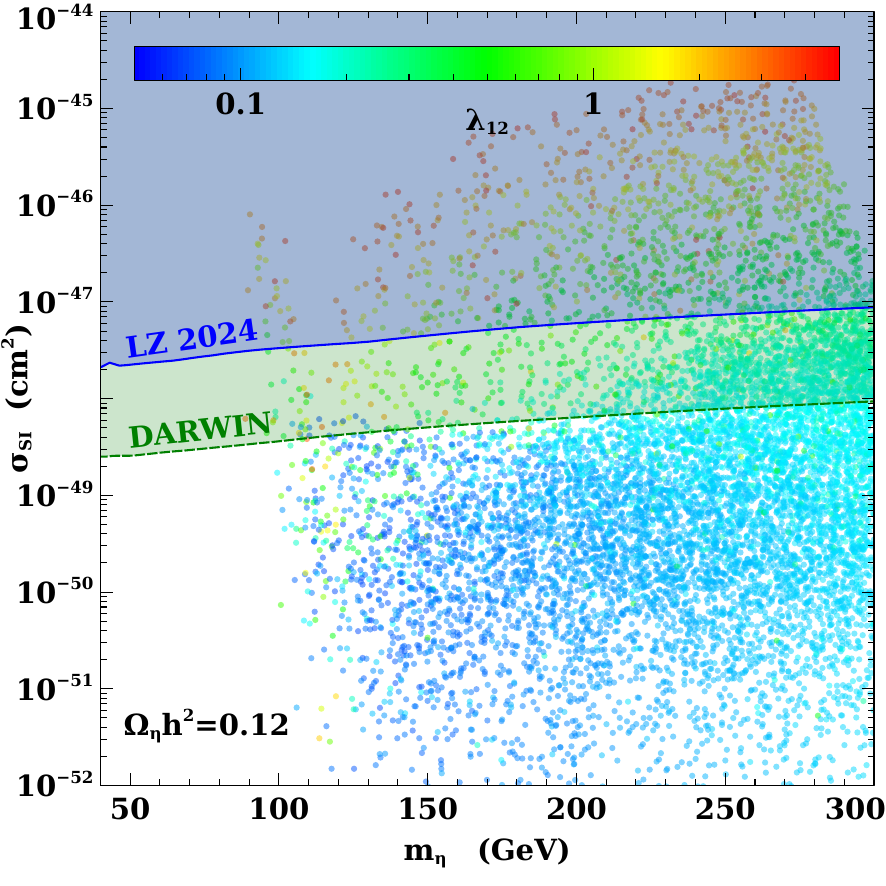}
    \caption{Left panel: Parameter space satisfying DM relic density for pure singlet scalar DM as well as the extension by a charged scalar. Right panel: Spin-independent one-loop DM-nucleon cross-section for the parameter space satisfying correct relic density. See text for details.}
    \label{fig:relic_density} 
\end{figure*}

\begin{figure}[h]
    \centering
    \includegraphics[width=0.9\linewidth]{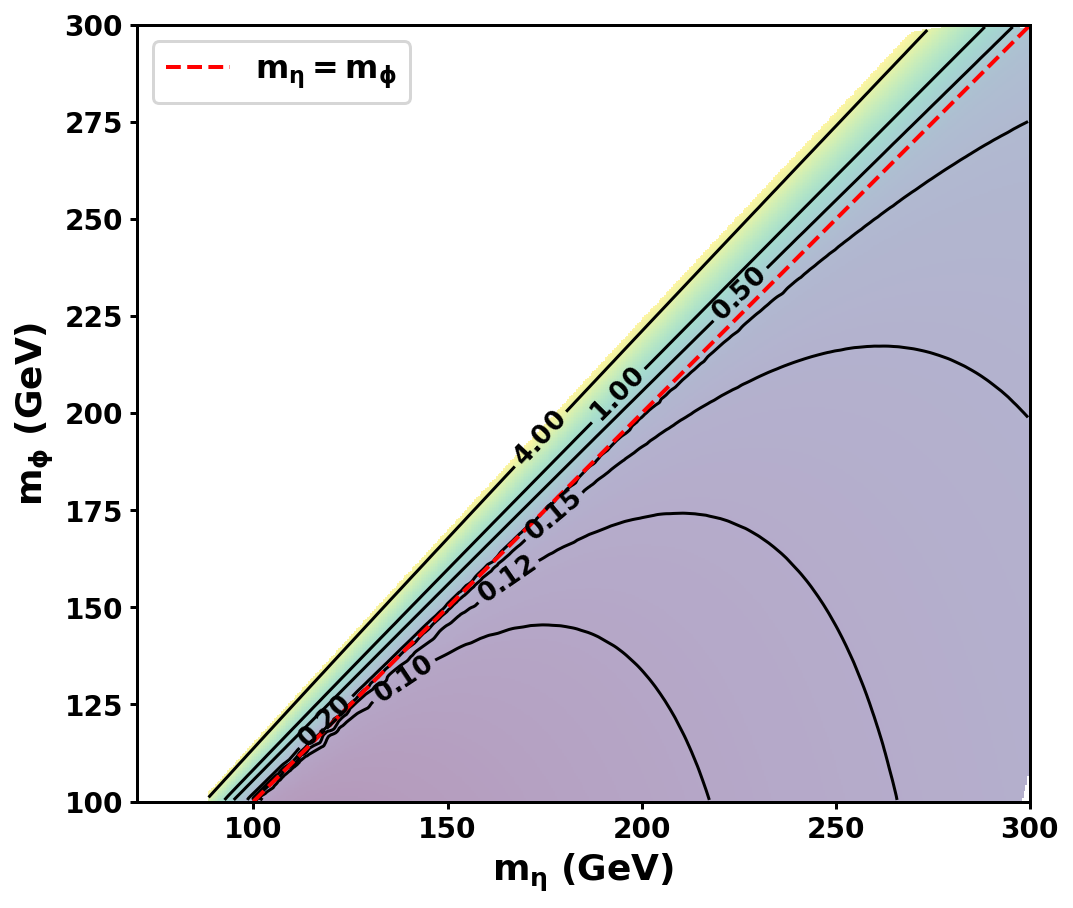}
    \caption{Parameter space satisfying the DM relic density, $\Omega_{\eta} {\rm h}^2=0.12$. The shaded color represents the value of $|\lambda_{12}|$ as it varies from 0 (violet) to 6 (yellow). The black curves represent contours of constant $|\lambda_{12}|$ corresponding to values \{0.1,\,0.15,\,0.2,\,1,\,4\}.}
    \label{fig:relic_density2}
\end{figure}

In addition to the bounds from relic abundance criteria, scalar singlet DM parameter space is also constrained from non-observation of DM-nucleon scattering at direct-detection experiments. The same Higgs portal coupling $\lambda_{13}$ via which scalar singlet DM annihilates also lead to elastic DM-nucleon scattering $\eta N \rightarrow \eta N$, where $N$ is a nucleon. The left panel of Fig.\,\ref{fig:relic_density} depicts the parameter space in the plane of Higgs portal coupling $\lambda_{13}$ and DM mass $m_\eta$, comparing the pure singlet scalar DM scenario with the charged-scalar portal extension discussed in this work. The shaded blue region is ruled out by spin-independent direct-detection limits from LUX-ZEPLIN (LZ) ~\cite{LZ:2024zvo} while the green-shaded region is the projected DARWIN \cite{DARWIN:2016hyl} sensitivity. The solid red contour shows the status of the minimal scalar singlet DM with the entire relic-satisfied parameter space in the sub-TeV ballpark being ruled out except the narrow tip around the Higgs resonance. The purple colored points correspond to the parameter space of charged scalar portal extension of scalar singlet DM consistent with relic density. While the Higgs portal coupling $\lambda_{13}$ is kept small to avoid large tree level DM-nucleon scattering, DM relic is primarily governed by charged scalar portal coupling $\lambda_{12}$. However, $\phi$ mediated diagrams at one-loop can contribute significantly to the spin-independent DM-nucleon cross-section due to the portal couplings $\lambda_{12}$ and $\lambda_{23}$. The details of the one-loop DM-nucleon cross-section is given in Appendix \ref{sec: dd_1loop}. The photon mediated diagram vanishes due to EM gauge symmetry, while the $Z$-mediated diagram is suppressed compared to the $h$-mediated diagram by a factor of $q^2/m_Z^2$, where $q$ is the momentum transfer. The spin-independent one-loop cross-section, $\sigma_{\rm{SI}}$ gets leading contribution from the $h$-mediated diagram. In the non-relativistic (NR) limit, we have
\begin{align}
\sigma^{\rm{NR}}_{\rm{SI}} &= \frac{\mu^2_{\eta p}f_N^2}{4\pi m_{\eta}^2m_h^4}~\lambda_{\eta\eta h}^2\,,
\end{align} 
where the one-loop dimensionless portal coupling $\lambda_{\eta\eta h}$ is given by 
\beq 
\lambda_{\eta\eta h}(\mu)=  \lambda_{13}(\mu)+\frac{\lambda_{12}(\mu)\lambda_{23}(\mu)}{16\pi^2}\ln\frac{m_{\phi}^2}{\mu^2}\,.
\label{eq:eeh}
\eeq

The right panel of Fig.\,\ref{fig:relic_density} shows the one-loop spin-independent DM-nucleon cross-section for the same relic-satisfied points of charged scalar portal DM shown on the left panel. The color code shows the size of DM coupling to charged scalar with larger $\mathcal{O}(1)$ values of $\lambda_{12}$ being ruled out by current direct-detection limits.

In the limit of small Higgs portal coupling $\lambda_{13}$ of DM, the relic abundance is primarily decided by the annihilation $\eta\eta\rightarrow\phi\phi$ with the cross-section given in Eq.\,\eqref{eq:etaphi}. The relic allowed parameter space is shown in Fig.\,\ref{fig:relic_density2} in $m_\phi-m_\eta$ plane where different solid contours correspond to a fixed value of $\lambda_{12}$. For large $\mathcal{O}(1)$ values of $\lambda_{12}$, DM remains in the kinematically forbidden regime $m_\eta < m_\phi$ \cite{DAgnolo:2015ujb, DAgnolo:2020mpt, Griest:1990kh}. However, this regime is ruled out by direct-detection bounds due to large one-loop DM-nucleon scattering, as shown on the right panel of Fig.\,\ref{fig:relic_density}.

\section{The effective potential}
\label{sec:3}
The thermal evolution of the \vev in the early Universe is governed by the effective potential which we write in terms of background fields $(\eta,\varphi,h)$. In the Arnold-Espinosa prescription \cite{Arnold:1992rz}, the one-loop daisy-improved effective potential is written as
\beq 
\label{eq: Veff}
V_{\rm{eff}} = V_0 + V_{\rm{CW}} + V_{\rm{CT}} + V_{1T} + V_{\rm{daisy}}\,.
\eeq 

The tree-level piece is
\begin{align}
V_0(\eta,\varphi,h)&=\frac{\mu_1^2}{2}\,\eta^2
+\frac{\mu_2^2}{2}\,\varphi^2
-\frac{\mu_3^2}{2}\,h^2\nn\\
&+\frac{\lambda_1}{4}\,\eta^4
+\frac{\lambda_2}{4}\,\varphi^4
+\frac{\lambda_3}{4}\,h^4\nn\\
&+\frac{\lambda_{12}}{4}\,\eta^2\varphi^2
+\frac{\lambda_{13}}{4}\,\eta^2 h^2
+\frac{\lambda_{23}}{4}\,h^2\varphi^2\,.
\end{align}

The one-loop correction is given by the Coleman-Weinberg potential \cite{Coleman:1973jx}. In the $\overline{\rm{MS}}$ scheme and in Landau gauge, the expression is
\beq
V_{\rm{CW}}(\Phi) = \frac{1}{64\pi^2}\sum_{i} (-1)^{f_i}  n_i m_i^4(\Phi)\left[\log\left(\frac{m_i^2(\Phi)}{\mu^2}\right)-c_i\right]\,,
\eeq
where $\Phi$ denotes the scalar fields $(\eta,\varphi,h)$, $m_i$ are the field-dependent masses of all species coupled to the scalars, $n_i$ are the corresponding degrees of freedom (DOF). $f_i$ is $+1$ for fermions and $0$ for bosons, and $\mu$ is the renormalization scale. The constant $c_i$ is $5/6$ for gauge bosons, and $3/2$ for other fields. The field-dependent masses are given in Appendix\,\ref{sec: masses}. The dof for ($\eta,\,\varphi,\,\xi,\,h,\,G_0,\,G^{\pm},\,W^{\pm}$,\,t) are $n_{i}=(1,\,1,\,1\,,1,\,1,\,2,\,6,\,12)$ respectively.

To ensure the masses of physical scalars, and \vevs, remain at their tree-level values, we add a finite counter-term potential, which has the same form as $V_0$
\begin{align}
V_{\rm{CT}}(\eta,\varphi,h)&=\frac{\delta\mu_1^2}{2}\,\eta^2
+\frac{\delta\mu_2^2}{2}\,\varphi^2
-\frac{\delta\mu_3^2}{2}\,h^2\nn\\
&+\frac{\delta\lambda_1}{4}\,\eta^4
+\frac{\delta\lambda_2}{4}\,\varphi^4
+\frac{\delta\lambda_3}{4}\,h^4\nn\\
&+\frac{\delta\lambda_{12}}{4}\,\eta^2\varphi^2
+\frac{\delta\lambda_{13}}{4}\,\eta^2 h^2
+\frac{\delta\lambda_{23}}{4}\,h^2\varphi^2\,.
\end{align}

Next, we include the one-loop finite temperature correction \cite{Dolan:1973qd,Quiros:1999jp}
\beq
    V_{1T}(\Phi,T) = \frac{T^4}{(2\pi)^2}\sum_i (-1)^{f_i} n_i J_{b/f}\left(\frac{m_i^2}{T^2}\right),
\eeq
where the functions $J_{b/f}$ are given by.
\beq
  J_{b/f}(x^2) = \int_0^{\infty} dy~ y^2 \log[1\mp e^{-\sqrt{y^2+x^2}}].
\eeq
Here $b (-), f (+)$ correspond to boson and fermion respectively.

The final piece in the effective potential is the daisy correction, obtained by resumming higher-loop daisy diagrams 
\beq
V_{\rm{daisy}} = -\frac{T}{12\pi}\sum_b n_b\Big((m_b^2+\Pi_b)^{3/2}-m_b^3\Big)\,
\eeq
where the bosonic thermal squared masses $\Pi_b$ are listed in Appendix\,\ref{sec: masses}. 

\subsection{Non-restoration of $U(1)_Y$ symmetry}
\label{sec: symmetry_nonrestoration}
Non-restoration of gauge symmetries in the standard model and beyond at high temperatures has been studied for a long time \cite{Weinberg:1974hy, Salomonson:1984rh, Dvali:1995cj, Gavela:1998ux, Cline:1999wi, Espinosa:2004pn, Patel:2013zla, Meade:2018saz, Baldes:2018nel, Matsedonskyi:2020mlz, Matsedonskyi:2021hti, Carena:2021onl, Bai:2021hfb, Chao:2021xqv, Lopez-Zurdo:2023xmr}. Such scenarios not only have additional discovery prospects due to their rich phenomenology but can also provide solutions to other puzzles in particle physics like the origin of dark matter and baryon asymmetry of the Universe. In this work, we explore the non-restoration of gauged hypercharge symmetry $U(1)_Y$ of the standard model at high temperatures through a non-zero \vev of $\varphi$. To see which regions of the parameter space allow for such symmetry non-restoration, we write the effective potential in the high temperature limit
\beq
V_{\rm{eff}} \approx V_0 + V_{1T}^{\rm{high}},
\eeq
where 
\beq \label{eq: 1_V1T_high}
    V_{1T}^{\rm{high}} \approx \frac{T^2}{24}\left(\sum_b n_b m_b^2 + \frac{1}{2}\sum_f n_f m_f^2\right)\, 
\eeq
is the high-temperature expansion of the thermal potential. Here $m_b$, $m_f$ are the field-dependent masses of bosons and fermions, respectively, and $n_b$, $n_f$ are the corresponding DOF. Thus,
\begin{align}
V_{\rm{eff}} &= V_0 + \frac{T^2}{2}(C_{\eta}^0 {\eta}^2+C_{\varphi}^0 {\varphi}^2+C_h^0 h^2)\,,\\
&=\frac{1}{2}\,(\mu_1^2+C^0_{\eta}T^2)\,\eta^2
+\frac{1}{2}\,(\mu_2^2+C^0_{\varphi}T^2)\,\varphi^2\nn\\
&+\frac{1}{2}\,(-\mu_3^2+C^0_hT^2)\,h^2+\frac{\lambda_1}{4}\,\eta^4
+\frac{\lambda_2}{4}\,\varphi^4\nn\\
&+\frac{\lambda_3}{4}\,h^4+\frac{\lambda_{12}}{4}\,\eta^2\varphi^2
+\frac{\lambda_{13}}{4}\,\eta^2 h^2
+\frac{\lambda_{23}}{4}\,h^2\varphi^2\,,\label{eq: highTVeff}
\end{align}
where the coefficients $C^0_i$ are given by
\begin{align}
    C_{\eta}^0 &= \frac{1}{12}(3\lambda_1+\lambda_{12}+2\lambda_{13})\,,\label{eq: C0_eta}\\
    C_{\varphi}^0 &= \frac{1}{24}\Big(6 g'^2 +8\lambda_2+\lambda_{12}+4\lambda_{23}\Big)\,,\label{eq: C0_varphi}\\
    C_{h}^0 &= \frac{1}{48}\Big(9g^2+3g'^2+12y_t^2+24\lambda_3+2\lambda_{13}+4\lambda_{23}\Big)\,.\label{eq: C0_h}
\end{align}

Next, we include the daisy potential in the high-$T$ limit. We denote the thermal masses as $\Pi_b(T) = D_b T^2$. In the high temperature limit, $m_b^2\ll\Pi_b$, the daisy term takes a simple form
\beq
V^{\rm{high}}_{\rm{daisy}} \approx -\frac{T^2}{8\pi}\sum_b\Big(n_b\sqrt{D_b} m_b^2\Big)\,.
\eeq

\begin{figure*}
    \centering
    \includegraphics[width=0.8\linewidth]{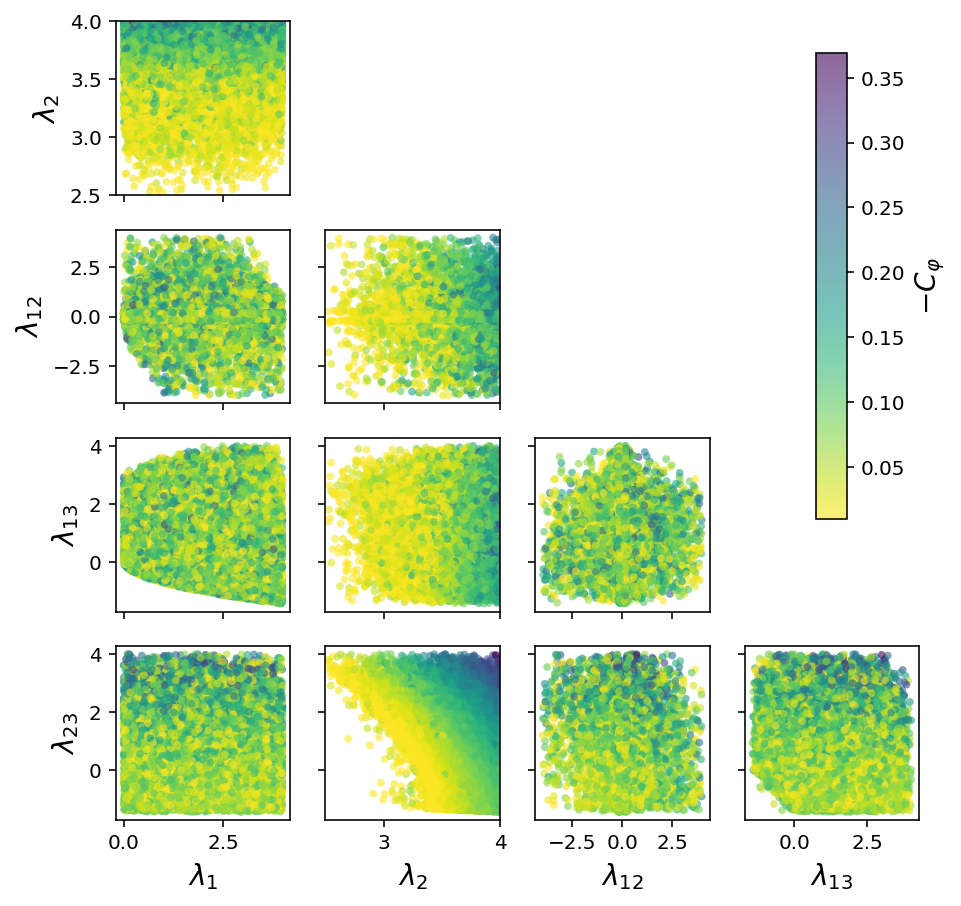}\hfill
    \caption{Scan over the couplings $\{\lambda_1,\,\lambda_2,\,\lambda_{12},\,\lambda_{13}\,,\lambda_{23}\}$, where $C_\varphi<0$, as required for $U(1)_Y$ symmetry non-restoration. The color code indicates the $C_\varphi$ value.}
    \label{fig:scan_cphi1} 
\end{figure*}

\begin{figure}
    \centering    
    \includegraphics[width=.9\linewidth]{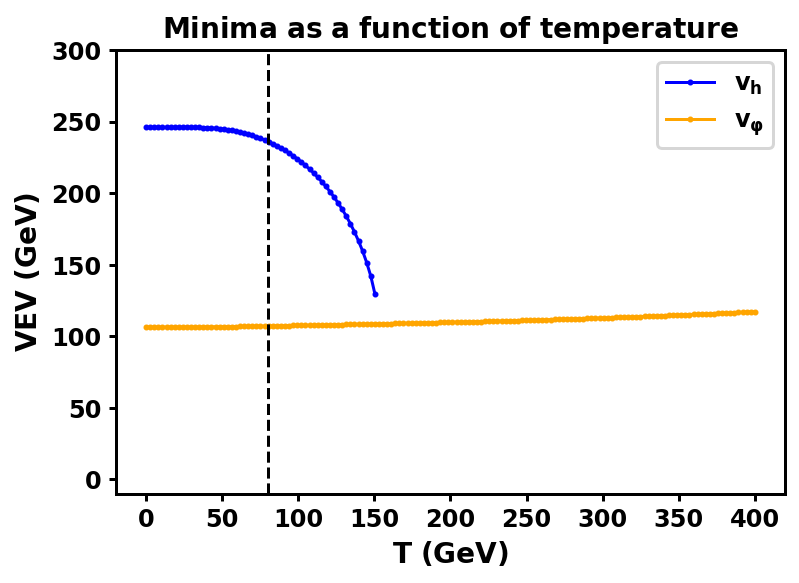}\hfill   
    \caption{Evolution of the two minima at $(0,v_h)$ and $(v_\varphi,0)$ with temperature for the benchmark point BP1. The value of $v_\varphi$ increases with temperature, indicating  symmetry non-restoration. The dashed black line indicates the critical temperature $T_c$ (see Sec.\,\ref{sec: fopt}).}
    \label{fig: phases}
\end{figure}

Including these daisy contributions in the high-temperature limit of the effective potential leads to
\beq
V_{\rm{eff}} \approx V_0 + V_{1T}^{\rm{high}}+V^{\rm{high}}_{\rm{daisy}}\,,
\eeq
where the daisy correction gives an additional contribution $\Delta C^0_i$ to the coefficients $C^0_i$ appearing in Eq.\eqref{eq: highTVeff}, such that $C^0_i\rightarrow C_i=C^0_{i}+\Delta C^0_{i}$. These shifts are given by
\begin{align}
    -4\pi\Delta C^0_{\eta} &=3\sqrt{C^0_{\eta}}\,\lambda_1+\sqrt{C^0_{\varphi}}\,\lambda_{12}+ 2\sqrt{C^0_{h}}\,\lambda_{13}\,,\\
    -4\pi\Delta C^0_{\varphi} &= \sqrt{\frac{19}{6}} \, g'^3+4 \sqrt{C^0_{\varphi}}\,\lambda_2+\frac{1}{2}\sqrt{C^0_{\eta}}\,\lambda_{12} \nn \\
    &+2\sqrt{C^0_{h}}\,\lambda_{23}\,, \label{eqcphi}\\
    -4\pi\Delta C^0_h &= \frac{\sqrt{66}}{8} g^3 + \frac{\sqrt{114}}{24} g'^3+ 6\sqrt{C^0_{h}}\,\lambda_3\nn\\
    &+\frac{1}{2}\sqrt{C^0_{\eta}}\,\lambda_{13}+\sqrt{C^0_{\varphi}}\,\lambda_{23}\,.
\end{align}

The sign of $C_i$ determines whether the effective potential at the origin is stabilized or destabilized along the `$i$' direction at high temperatures. When $C_{\varphi}<0$, $\varphi$ develops a non-zero \vev at $T\gg v$, breaking the $U(1)_Y$ symmetry in the early universe. We are thus interested in the scenario 
$$ C_{\eta}\,,~C_{h}\geq 0\,,\quad C_{\varphi}< 0\,. $$

To find the parameter space compatible with the above conditions along with the copositivity and perturbativity criteria, we perform a scan over the following parameters:
\beq
\{\lambda_1,\,\lambda_2,\,\lambda_{12},\,\lambda_{13},\,\lambda_{23}\}. 
\eeq
The scan indicates that coefficients $C^0_i$ always stay positive and that the daisy corrections are necessary to get $C_{\varphi}<0$. Fig.\,\ref{fig:scan_cphi1} shows the parameter space of the model consistent with the requirement $C_{\varphi}<0$, necessary to achieve $U(1)_Y$ non-restoration in the early Universe. As this figure indicates, $\mathcal{O}(1)$ values of the relevant quartic couplings lead to larger negative values of $C_{\varphi}$ which is also clear from Eq. \eqref{eqcphi}. In the numerical scan, we keep the charged singlet scalar mass above $100$ GeV, typical lower limit from LEP data \cite{Das:2025oww}. Fig.\,\ref{fig: phases} shows how the $(v_\varphi,0)$ minimum moves away from the origin with temperature for the benchmark point BP1 listed in Table\,\ref{table: bp}, indicating the non-restoration of $U(1)_Y$ symmetry.

\subsection{First-order Phase Transition}
\label{sec: fopt}
As $\eta$ never develops a non-zero VEV in the scenario under consideration, we set $\eta = 0$ in the effective potential and analyse the possible phase transitions in the $\varphi-h$ plane. Let us define $T_Y$ as the temperature of $U(1)_Y$ breaking via $\varphi$ \vev, and $T_{\rm{EW}}$ as the temperature below which Higgs gets a \vev. 
If both \vevs are non-zero in a certain temperature range, i.e. $T_Y<T_{\rm{EW}}$, there is a possibility of having two coexisting minima of the effective potential; one at $(0,v_h)$, and another at $(v_\varphi, 0)$, separated by a barrier, leading to a FOPT. Using the full finite-temperature effective potential given in Eq.\,\eqref{eq: Veff}, we calculate the critical temperature $T_c$ at which the two minima become degenerate. We use the \texttt{CosmoTransitions} package \cite{Wainwright:2011kj} to track the phases of the effective potential. The field jump during FOPT can be quantified by 
\beq
v_c \equiv \left.\sqrt{v^2_h+v^2_\varphi}\right\vert_{T=T_c}\,.
\eeq 
The ratio $v_c/T_c$ is a measure of the strength of FOPT, where strong FOPTs satisfy the criterion
\beq
\frac{v_c}{T_c} > 1\,. 
\eeq

\begin{figure*}
    \centering    
    \includegraphics[width=1.0\linewidth]{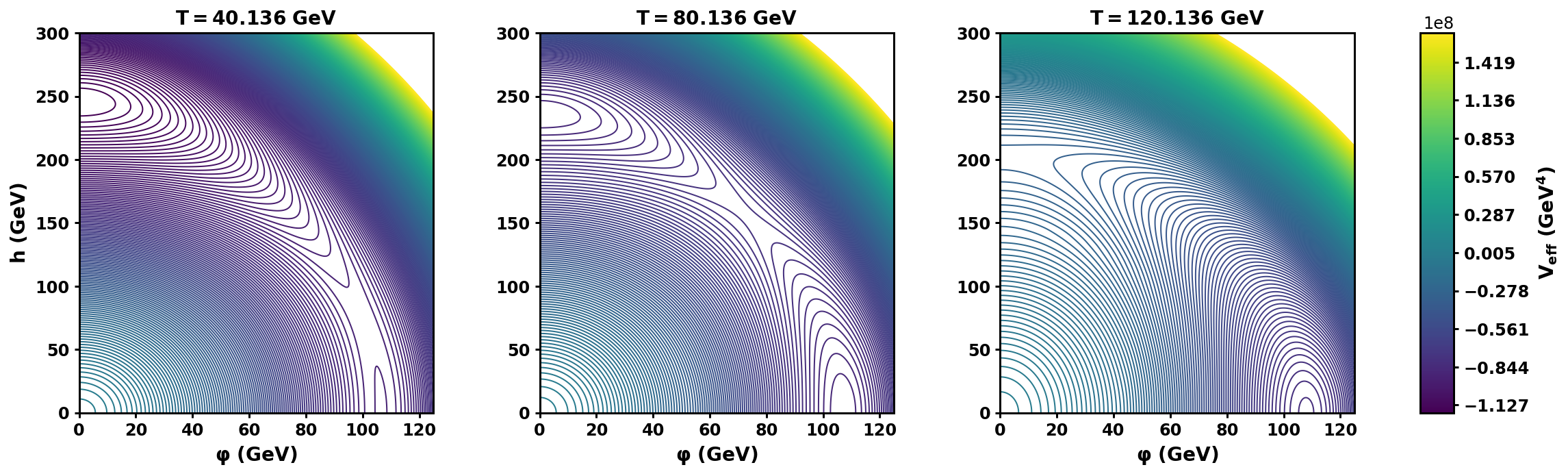}\hfill   
    \caption{
    Evolution  of the effective potential with temperature for the benchmark point BP1 given in Table \ref{table: bp}. 
    In a finite temperature range, the minima at $(0,v_h)$ and $( v_\varphi,0)$ coexist, leading to FOPT. The plots correspond to three different temperatures: $T<T_c$ (left panel), $T=T_c$ (middle panel), and $T>T_c$ (right panel). The FOPT progresses from right to left.}

    \label{fig:contour_plots}
\end{figure*}

\begin{figure*}
    \centering    \includegraphics[width=0.8\linewidth]{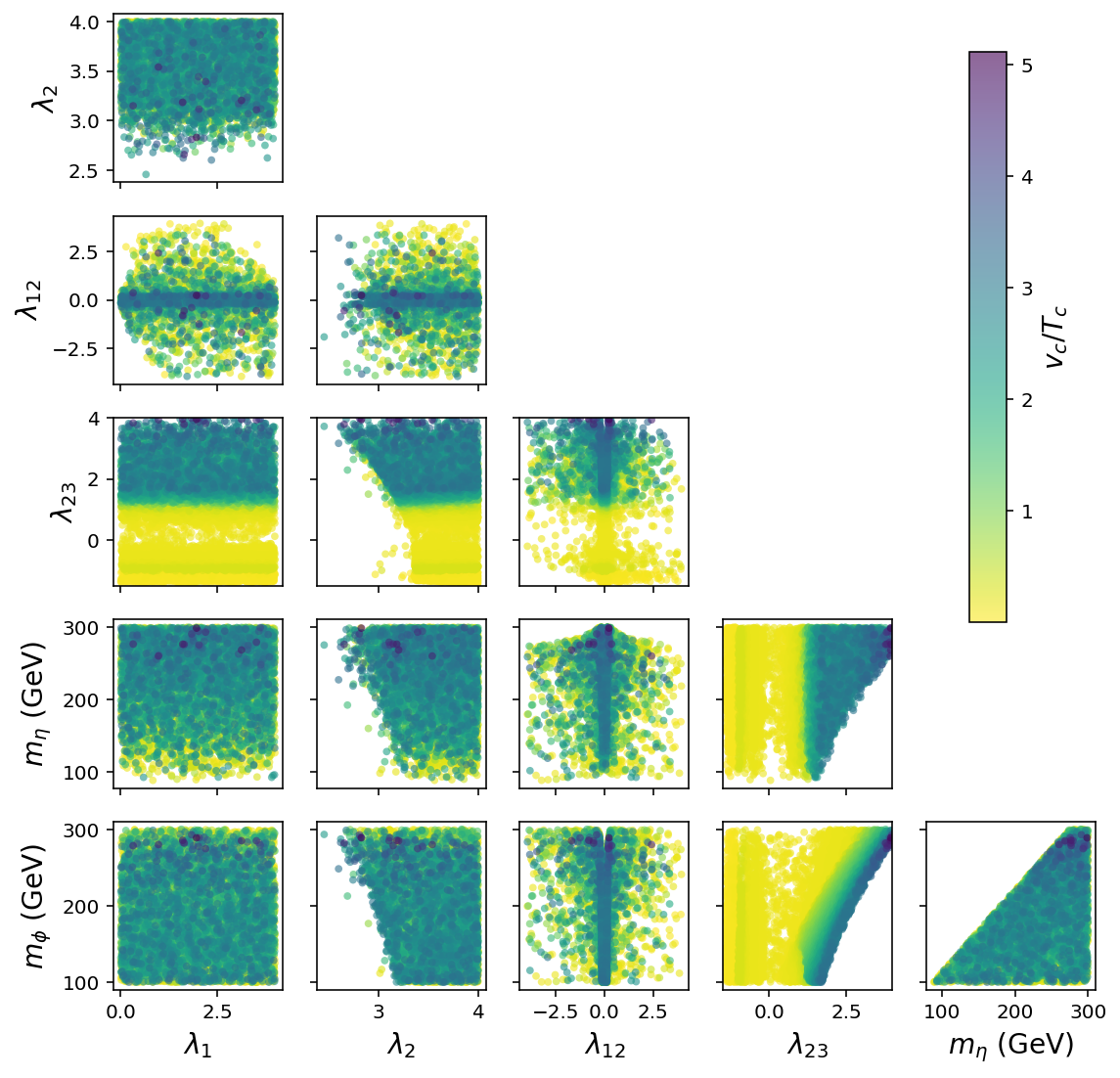}\hfill   
    \caption{Parameter space of the model consistent with FOPT during the transition from the $U(1)_Y$ non-restored minimum to the EW minimum, with the color bar indicating the ratio $v_c/T_c$.}
    \label{fig:scan_cphi} 
\end{figure*}

Below $T_c$ the FOPT proceeds via tunnelling of the false vacuum $(\langle\varphi\rangle,\langle h\rangle) = (v_\varphi,0)$ to the true vacuum $(\langle\varphi\rangle,\langle h\rangle) = (0,v_h)$.  The rate of tunnelling is estimated by calculating the bounce action $S_3$ using the prescription in \cite{Linde:1980tt}. The nucleation temperature $T_n$ is then calculated by comparing the tunnelling rate $\Gamma$ with the Hubble expansion rate of the Universe $\Gamma (T_n) = \mathcal{H}^4(T_n) \equiv \mathcal{H}^4_*$. Fig.\,\ref{fig:contour_plots} shows finite-temperature potential for the benchmark point BP1 at $T < T_c$ (left), $T=T_c$ (middle), $T > T_c$ (right) depicting the first-order phase transition from charge-breaking minima to the standard Higgs minima as the temperature cools. Fig.\,\ref{fig:scan_cphi} shows the parameter space of the model consistent with such a FOPT with the color bar indicating the ratio $v_c/T_c$. Points satisfying $v_c/T_c>1$ have a large value of the portal coupling $\lambda_{23}$, ($\lambda_{23}\gtrsim 1.5$). According to Eq.\,\ref{eq: mphi}, this implies $\mu_2^2<0$ whenever $m_\phi\lesssim 213\,$ GeV. Hence at tree-level, a local minimum along the $\varphi$ direction assists in enhancing the strength of FOPT.

One of the appealing features of a strong FOPT in the early Universe is the generation of stochastic gravitational wave background due to bubble collisions~\cite{Turner:1990rc,Kosowsky:1991ua,Kosowsky:1992rz,Kosowsky:1992vn,Turner:1992tz}, the sound wave of the plasma~\cite{Hindmarsh:2013xza,Giblin:2014qia,Hindmarsh:2015qta,Hindmarsh:2017gnf} and the turbulence of the plasma~\cite{Kamionkowski:1993fg,Kosowsky:2001xp,Caprini:2006jb,Gogoberidze:2007an,Caprini:2009yp,Niksa:2018ofa}. While the peak frequency and peak amplitude of such GW spectrum depend upon specific FOPT related parameters, the exact nature of the spectrum is determined by numerical simulations. One of the key parameters for GW estimates namely, the latent heat released in the FOPT is measured by the parameter $\alpha$, defined as
\beq 
\alpha = \frac{1}{\rho_{\rm{rad}}}\left.\left(\Delta V - \frac{T}{4}\frac{\partial\Delta V}{\partial T}\right)\right\vert_{T=T_*}\,,
\eeq
where $\rho_{\rm{rad}} = \frac{\pi^2}{30}g_*(T) T^4$ is the radiation energy density of the thermal bath, $g_* \approx 110$ is the number of relativistic degrees of freedom, $T_*$ is the temperature at which GW production is dominant and $\Delta V$ is the energy difference in true and false vacua. For this work we take $T_*=T_n$. Another key parameter is the rate of the phase transition, expressed in terms of the parameter $\beta$, defined via the expression
\beq 
\frac{\beta}{\mathcal{H}_*} = -T\frac{d}{dT}\left.\left(\frac{S_3}{T}\right)\right\vert_{T=T_*}
\eeq
The bubble wall velocity $v_w$ is estimated from the Jouguet velocity \cite{Kamionkowski:1993fg, Steinhardt:1981ct, Espinosa:2010hh}
$$v_J = \frac{1/\sqrt{3} + \sqrt{\alpha^2 + 2\alpha/3}}{1+\alpha}$$
following the prescription given in \cite{Lewicki:2021pgr}. This leads to the bubble wall velocity $v_w$ as \cite{Lewicki:2021pgr}
\begin{equation}
    v_w = 
    \begin{cases}
    \sqrt{\frac{\Delta V}{\alpha \rho_{\rm rad}}} & \text{if} \,\, \sqrt{\frac{\Delta V}{\alpha \rho_{\rm rad}}} < v_J \\
    1 & \text{if}  \,\, \sqrt{\frac{\Delta V}{\alpha \rho_{\rm rad}}} \geq v_J. \\
    \end{cases}
\end{equation}
For the benchmark points given in Table\,\ref{table: bp}, $\alpha\gtrsim\mathcal{O}(0.1)$, we can safely take $v_w=1$.
 
\begin{table}[h] 
\begin{center}
\vspace{0.5 cm}
\begin{tabular}{|c|c|c|c|c|c|}
\hline
Parameters & BP1   & BP2   & BP3    & BP4 & BP5\\ 
\hline
$m_{\eta}$ (GeV) & 206.313 & 249.776 & 161.453 & 161.866 & 248.775 \\  
$m_{\phi}$ (GeV) & 109.127 & 104.950   & 100.230  & 110.740 & 113.048 \\ 
$\lambda_1$ & 0.074 & 1.266  & 0.209 & 0.187 & 0.870   \\  
$\lambda_2$ & 3.533 & 3.523  & 3.654 & 3.426 & 3.767  \\ 
$\lambda_{12}$ & -0.095 & -0.112  & 0.076 & 0.079 & -0.112  \\  
$\lambda_{13}$  & -0.001  & -0.001 & -0.001 & -0.001 & -0.001  \\ 
$\lambda_{23}$ & 1.738 & 1.708  & 1.702 & 1.678 & 1.752   \\  
\hline
$v_c/T_c$ & 3.236  & 3.257 & 3.275 & 2.932 & 2.891   \\
$\alpha$ & 0.886   & 0.384   & 0.207 & 0.142 & 0.118   \\ 
$\beta/\mathcal{H}_*$ & 220.380 & 366.500   & 699.270 & 661.880 & 891.450 
   \\  
$T_c$ (GeV) & 80.136 & 79.685   & 79.284 & 87.045 & 88.030    \\
$T_n$ (GeV) & 32.500 & 41.233   & 49.826 & 59.025 & 63.219   \\
$\Lambda_{\rm{NP}}$ (TeV) & 1.222 & 1.231 & 1.138 & 1.312 & 1.065\\
\hline
\end{tabular}
\end{center}
\caption{Benchmark points with early hypercharge breaking due to symmetry-non restoration, which exhibit FOPT, satisfy DM relic density, and evade direct-detection bounds.}
\label{table: bp} 
\end{table}

\section{Detection Prospects}
\label{sec:det}

\begin{figure}
    \centering
    \includegraphics[width=0.95\linewidth]{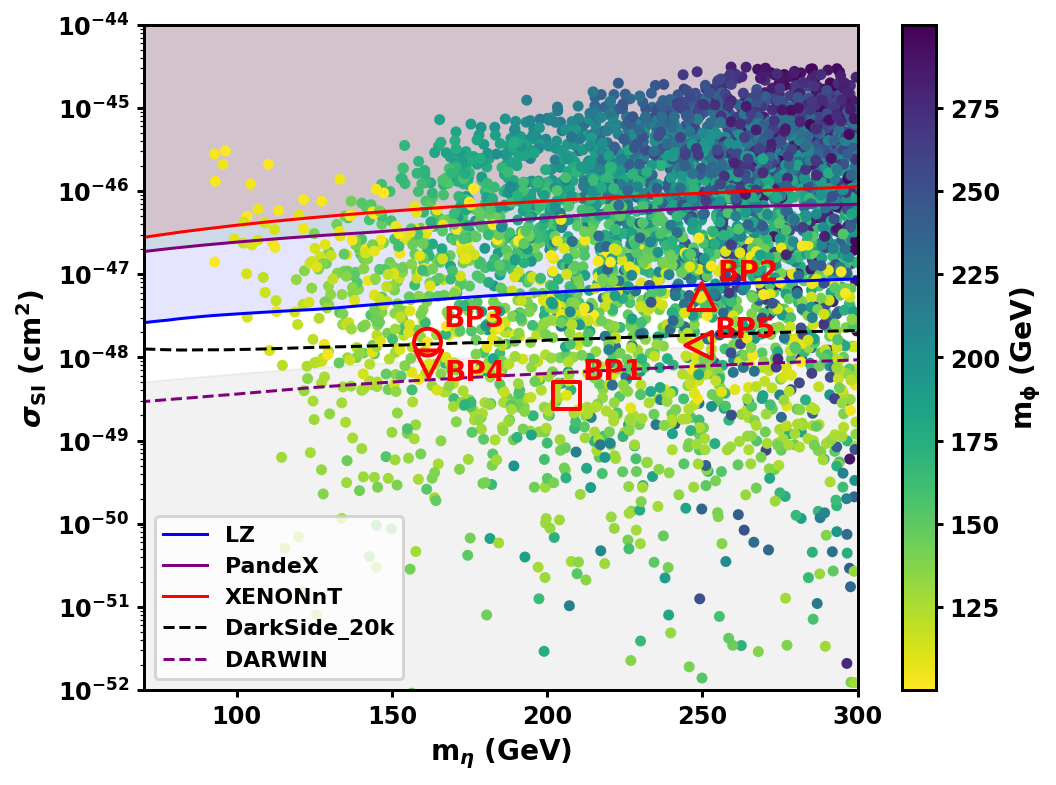}
    \caption{Variation in spin-independent DM-nucleon cross-section generated at one-loop due to charged scalar portal couplings with DM mass for the parameter space consistent with DM relic and strong first-order FOPT $v_c/T_c>1$. Color code indicates the charged scalar mass while solid and dashed contours correspond to current and future limits from direct-detection experiments. The highlighted points correspond to the benchmark points given in Table \ref{table: bp}.}
    \label{fig:lambda_max}
\end{figure}

As discussed above, the charged singlet scalar not only assists in reviving the singlet DM parameter space, but also keeps the scenario verifiable via DM-nucleon scattering as well as stochastic GWs emitted from the FOPT in the early Universe. Fig.\,\ref{fig:lambda_max} shows the one-loop spin-independent DM-nucleon scattering cross-section as a function of DM mass with the color code indicating the mass of the charged scalar. The solid contours correspond to upper limits from LUX-ZEPLIN (LZ) ~\cite{LZ:2024zvo}, XENONnT~\cite{XENON:2025vwd}, PandaX-4T~\cite{PandaX:2024qfu} respectively. The dashed contours correspond to sensitivities of next-generation experiments DarkSide$\_$20k \cite{DarkSide-20k:2017zyg}, DARWIN \cite{DARWIN:2016hyl} respectively. The points shown in this figure satisfy the criteria of correct DM abundance and a strong FOPT. A large part of the parameter space, particularly for larger $m_\phi$ is already ruled out by direct-detection limits on DM-nucleon cross-section while some part of it can be probed in near future. While effective one-loop DM-Higgs coupling has only logarithmic dependence on $m_\phi$ as shown in Eq. \eqref{eq:eeh}, increasing $m_\phi$ also increases $\lambda_{23}$ (cf. field-dependent masses in Appendix\,\ref{sec: masses}), leading to a more dominant effect. The benchmark points in Table \ref{table: bp} are also highlighted in the figure. They are allowed by the current direct-detection data but remain within reach of the next-generation experiments. 

\begin{figure}
    \centering
    \includegraphics[width=0.9\linewidth]{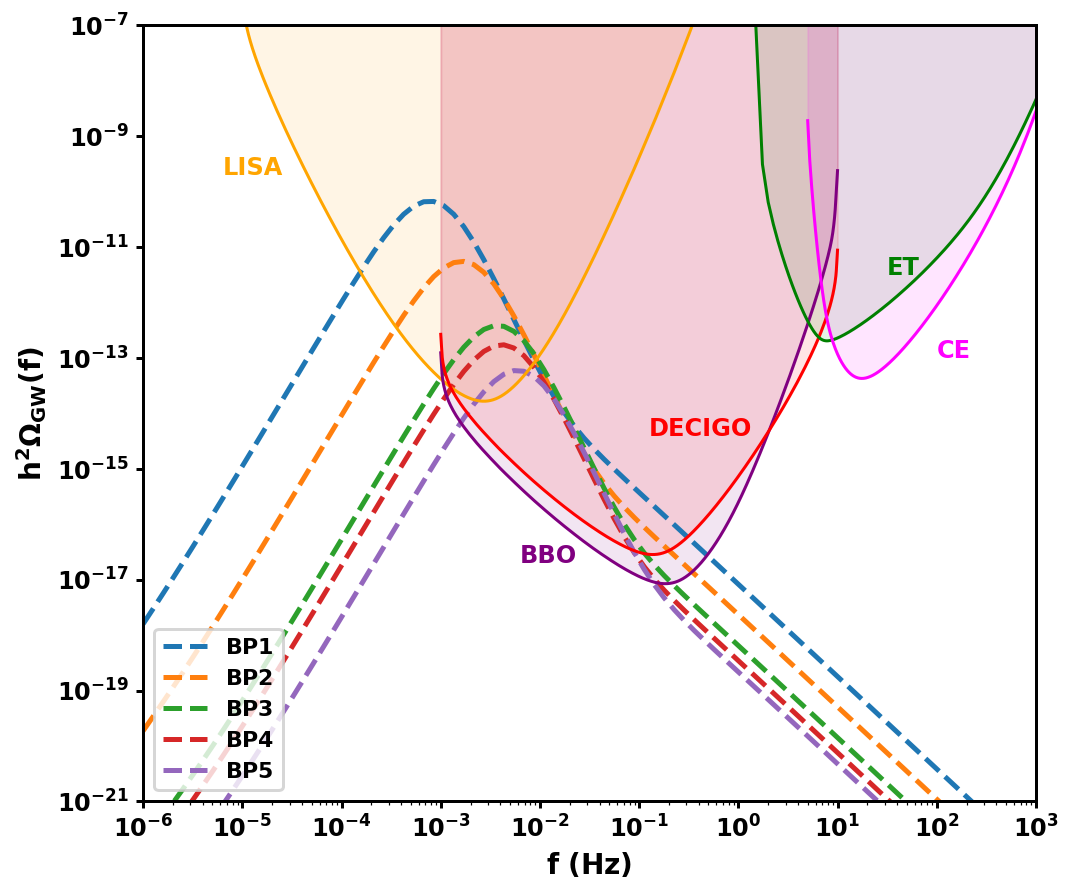}
    \caption{GW spectrum due from FOPT for the benchmark points listed in Table \ref{table: bp}. The power-law sensitivity curves for the GW detectors are taken from \cite{Schmitz:2020syl}.}
    \label{fig:gw}
\end{figure}

Fig.\,\ref{fig:gw} shows the GW spectra for the benchmark points in Table \ref{table: bp}. The FOPT-generated GW spectrum in this figure corresponds to the total GW spectrum obtained by adding the individual contributions from bubble collisions, sound waves and turbulence. The projected sensitivities of future GW observatories, including LISA~\cite{2017arXiv170200786A}, BBO~\cite{Yagi:2011wg}, DECIGO~\cite{Seto:2001qf,Kawamura:2006up,Yagi:2011wg}, ET~\cite{Punturo_2010} and CE~\cite{Hall:2022dik} are shown as shaded contours. Clearly, the peak GW frequencies for all the benchmark points remain within reach of LISA while BBO and DECIGO can also probe some of them. It should be noted that the requirement of a strong FOPT also forces some of the quartic couplings to assume large $\mathcal{O}(1)$ values. While they are perturbative at a chosen scale, they receive large corrections under renormalization group evolution (RGE) leading to non-perturbative regime at a higher scale $\Lambda_{\rm NP}$. Thus the perturbative treatment is no longer valid at a scale $\geq \Lambda_{\rm NP}$, requiring new strong dynamics or UV completions. 

While we have focussed only on the scalar portal couplings of the charged scalar, it can also have off-diagonal coupling to the leptons by virtue of Yukawa interaction $y_{\alpha \beta} \overline{\ell^c_\alpha} \ell_\beta \phi$ where $\ell$ denotes the SM lepton doublet. This allows the possibility of generating light neutrino masses via radiative seesaw \cite{Felkl:2021qdn}. They can also lead to the generation of lighter charged lepton masses at one-loop level \cite{Adhikari:2021yvx}. Such Yukawa interactions allow the singlet scalar to be probed at colliders via production and decay into leptonic final states \cite{Cao:2017ffm, Alcaide:2019kdr, Yue:2024bsk, Crivellin:2020klg, Das:2025oww}. More specifically, the charged singlet scalar can leave to observable lepton flavor universality violation \cite{Crivellin:2020klg} both at tree and radiative level. The scalar portal coupling of charged scalar with the SM Higgs can also lead to observable deviation of Higgs to diphoton rate from the one in the SM \cite{ATLAS:2022tnm, CMS:2021kom}. 

\section{Conclusion}
\label{sec:conc}
We have studied a minimal extension of the singlet scalar DM by a charged singlet scalar. Given the tightly constrained parameter space of the minimal singlet scalar DM, the singlet scalar adds new DM annihilation channels while keeping the direct-detection rates within limits due to one-loop suppression. As LEP bounds typically rules out singlet charged scalar with mass $\lesssim 100$ GeV \cite{Das:2025oww}, it also puts a lower bound on DM mass below which the charged scalar portal remains ineffective. We show the revival of singlet DM parameter space beyond $\sim 100$ GeV for a wide range of parameter space. Even though the Higgs mediated direct-detection rates remain suppressed, the charged scalar portal can lead to one-loop DM-nucleon scattering within current and future experiment's reach. Interestingly, the same predictive parameter space of charged scalar portal DM also leads to hypercharge symmetry non-restoration in the early Universe due to a non-zero \vev acquired by the charged scalar. A subset of the parameter space is also consistent with a first-order phase transition which takes the Universe to the standard Higgs minima, consistent with the conservation of electromagnetic charge. This keeps a part of the parameter space within reach of future GW experiments. However, this also forces some of the quartic couplings to have large $\mathcal{O}(1)$ values requiring UV completions at a scale $\gtrsim \mathcal{O}(1)$ TeV. Allowing charged scalar coupling to the leptons bring additional phenomenology and discovery prospects in terms of generating light neutrino mass, matter-antimatter asymmetry via leptogenesis and collider signatures. We leave these phenomenological studies as well as UV complete model building to future works.

\acknowledgments
The works of D.B. and I.S. are supported by the Science and Engineering Research Board (SERB), Government of India grant CRG/2022/000603. D.R. is supported by the ANRF-NPDF fellowship with scheme no. PDF/2025/003379.

\appendix

\section{One-loop contributions to $\sigma_{\rm{SI}}$}
\label{sec: dd_1loop}

\begin{figure}[h]
    \centering
    \includegraphics[width=0.95\linewidth]{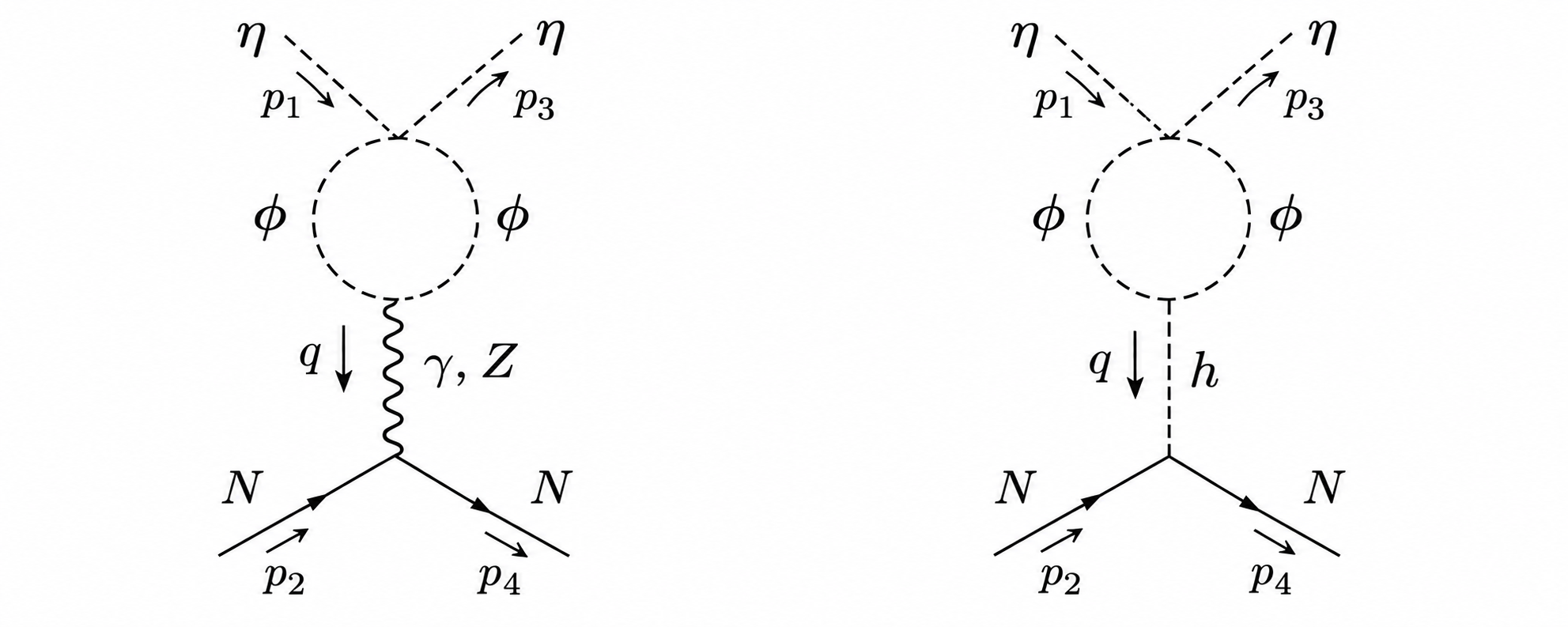}
    \caption{One-loop diagrams contributing to the direct-detection cross-section. Left panel: Gauge-mediated digrams, right panel: Higgs-mediated diagram.}
    \label{fig: feynman}
\end{figure}

The one-loop diagrams contribution to the spin-independent direct-detection cross-section are given in Fig.\,\ref{fig: feynman}.
First, we consider the gauge boson mediated diagrams shown in the left panel, which are absent at tree-level. For the photon ($\gamma$) mediated diagram, the scattering amplitude has the form
\beq
i\mathcal{M}^{(1)}_{\gamma} = -i\Gamma^{\mu}_{\eta\eta\gamma}\times \frac{-i\eta_{\mu\nu}}{q^2}\times J^{\nu}_N\,,
\eeq
where $q$ is the momentum transfer, $J^{\nu}_N$ is the nucleon EM current. $\Gamma^{\mu}_{\eta\eta\gamma}$ is the effective one-loop $\eta\eta\gamma$ vertex, given by
\begin{align}
-i\Gamma^{\mu}_{\eta\eta\gamma} &= -i \lambda_{12}\nn\\
&\times\int \frac{d^dk}{(2\pi)^d} \frac{-ie q^{\mu}}{(k^2-m_{\phi}^2)((q-k)^2-m_{\phi}^2)}\,\\
&= -i\lambda_{12}\int_0^1 dx \int \frac{d^d l}{(2\pi)^2} \frac{-ieq^{\mu}}{(l^2-\Delta(q^2,m_{\phi}^2))^2}\, \nonumber \\
&\equiv q^{\mu} F(q^2,m_{\phi}^2)\,,\label{eq: Gamma_photon}
\end{align}
where in the second step, we have used Feynman parametrization with 
\beq
l = k-q(1-x),\quad \Delta(q^2,m_{\phi}^2) = m_{\phi}^2-x(1-x)q^2.
\eeq
Using Eq. \eqref{eq: Gamma_photon} the matrix element is
\beq
i\mathcal{M}^{(1)}_{\gamma} = -i\frac{F(q^2,m_{\phi}^2)}{q^2}~q_{\mu}(J^{{\rm{EM}}}_N)^\mu\,.
\eeq
EM current conservation gives $q_{\mu} (J^{{\rm{EM}}}_N)^\mu=0$, implying that the matrix element vanishes, i.e. $\mathcal{M}^{(1)}_{\gamma}=0$\,.\\

The $Z$-mediated diagram can be evaluated in a similar manner
\begin{align}
i\mathcal{M}^{(1)}_{Z} &= -i\Gamma^{\mu}_{\eta\eta Z}\times \frac{-i(\eta_{\mu\nu}-\frac{q_{\mu}q_{\nu}}{m_Z^2})}{q^2-m_Z^2}\times J^{{\rm{Weak}}~\nu}_N\\
&= q^{\mu}F'(q^2,m_{\phi}^2)\frac{-i(\eta_{\mu\nu}-\frac{q_{\mu}q_{\nu}}{m_Z^2})}{q^2-m_Z^2}\times J^{{\rm{Weak}}~\nu}_N\\
&= i\frac{F'(q^2,m_{\phi}^2)}{m_Z^2}\times q_{\mu}J^{{\rm{Weak}}~\mu}_N
\end{align}
where $J^{{\rm{Weak}}~\mu}_N$ is the nucleon weak current, and the effective one-loop $\eta\eta Z$ vertex is of the form $-i\Gamma^{\mu}_{\eta\eta Z} = q^{\mu}F'(q^2,m_{\phi}^2)$. Unlike the case of the nucleon EM current, $q_{\mu}J^{{\rm{Weak}}~\mu}_N \neq 0$, and hence the matrix element does not vanish. However, it is suppressed by a factor of $q^2/m_Z^2$ in the small $q$ limit\footnote{Typically, $q\sim 50\,$keV in direct-detection experiments} and therefore is negligible compared to the Higgs mediated diagram shown in the right panel of Fig.\,\ref{fig: feynman}, which we evaluate next. The scattering amplitude is
\begin{align}
i\mathcal{M}^{\rm{full}}_{h} &= (-ig_{\eta\eta h})\times \frac{i}{q^2-m_h^2}\nn\\
&\times \langle N(p_4)|\left(-i\frac{m_Q}{v}\right) \bar{Q}Q|N(p_2)\rangle\,
\end{align}
where Q denotes the quark fields contributing to the scalar nucleon matrix element. Since
\begin{align}
\langle N|\frac{m_Q}{v} \bar{Q}Q|N\rangle &= g_{hNN}\,\bar{u}(p_4)u(p_2)\,,\nn\\
g_{hNN}&=\frac{f_N m_N}{v}\,, 
\end{align}
where $f_N\approx 0.3$ and the nucleon mass, $m_N\approx 1\,$GeV, and $u$ is the nucleon spinor, we have
\begin{align}
i\mathcal{M}^{\rm{full}}_{h} 
&= -i \frac{g_{\eta\eta h}\,g_{hNN}}{q^2-m_h^2}\bar{u}(p_4)u(p_2)\nn\\
&\approx i \frac{g_{\eta\eta h}\,g_{hNN}}{m_h^2}\bar{u}(p_4)u(p_2)\quad ({\text{for $q^2\ll m_h^2$}})\,.
\end{align}
The effective $\eta\eta h$ coupling is
\beq
g_{\eta\eta h}(\mu) =g^{(0)}_{\eta\eta h}(\mu)+g^{(1)}_{\eta\eta h}(\mu)\,,
\eeq 
where $g^{(0)}_{\eta\eta h}$ and $g^{(1)}_{\eta\eta h}$ is the tree-level and the one-loop contribution respectively
\begin{align}
    g^{(0)}_{\eta\eta h} &= \lambda_{13} v\,,\\
    g^{(1)}_{\eta\eta h} &=i\int \frac{d^dk}{(2\pi)^d} \frac{ \lambda_{12}\lambda_{23}v}{(k^2-m_{\phi}^2)((q-k)^2-m_{\phi}^2)} \\
    &= - \frac{\lambda_{12}\lambda_{23} v}{16\pi^2}B_0(q^2,m_{\phi}^2,m_{\phi}^2)\,.
\end{align}
Here $B_0(q^2,m_1^2,m_2^2)$ denotes the standard scalar bubble integral (scalar two-point Passarino-Veltmann function \cite{Passarino:1978jh}). In the limit $q^2\ll m_\phi^2$, in $\overline{\rm {MS}}$ scheme,
\beq
B_0(0,m_{\phi}^2,m_{\phi}^2) = -\ln\frac{m_{\phi}^2}{\mu^2}
\eeq
Defining a dimensionless portal coupling 
\beq 
\lambda_{\eta\eta h}\equiv\frac{g_{\eta\eta h}}{v}\,,
\eeq
we have,
\beq 
\lambda_{\eta\eta h}(\mu)=  \lambda_{13}(\mu)+\frac{\lambda_{12}(\mu)\lambda_{23}(\mu)}{16\pi^2}\ln\frac{m_{\phi}^2}{\mu^2}\,.
\eeq

Below $\mu=m_h$, the Higgs field is integrated out, and the DM-nucleon interaction is described by the  EFT operator
\beq
\mathcal{L}_{\rm{EFT}}\supset \frac{1}{2}C_N\eta^2 \bar{N}N
\eeq 
Scattering amplitude is calculated as
\begin{align}
i\mathcal{M}^{\rm{EFT}}_{h} &= i C_N~ \bar{u}(p_4)u(p_2)\,.
\end{align}


Matching the full theory with EFT at $\mu=m_h$ yields
\begin{align}
C_N &= \frac{g_{\eta\eta h}(m_h) g_{hNN}}{m_h^2}\nn\\
&=\frac{\lambda_{\eta h}(m_h) f_N\, m_N}{m_h^2}
\end{align}
Below $\mu=m_h$, $C_N$ remains frozen and thus its value at the matching scale can be used to evaluate the spin-independent (SI) cross-section. Using the fact that $\bar{u}u\approx 2m_N$ in the non-relativistic (NR) limit, we have
\begin{align}
\sigma^{\rm{NR}}_{\rm{SI}} &= \frac{\mu^2_{\eta N}}{16\pi m_{\eta}^2 m_N^2}|\mathcal{M}_h|^2\\
&=\frac{\mu^2_{\eta N}}{4\pi m_{\eta}^2 } C_N^2 \\
&=\frac{\mu^2_{\eta p}f_N^2}{4\pi m_{\eta}^2m_h^4}~\lambda_{\eta\eta h}^2.
\end{align} 

\section{Field-dependent masses}
\label{sec: masses}
The field-dependent scalar mass matrix is 
\begin{align}
m_{\eta}^2 &=  3\lambda_1\eta^2+\half(\lambda_{12}\varphi^2+\lambda_{13}h^2)+\mu_1^2\,,\\
m_{\varphi}^2 &= 3\lambda_2\varphi^2+\half(\lambda_{12}\eta^2+\lambda_{23}h^2)+\mu_2^2\,,\\
m_{\xi}^2 &= \lambda_2\varphi^2+\half(\lambda_{12}\eta^2+\lambda_{23}h^2)+\mu_2^2\,,\\
m_h^2 &= 3\lambda_3 h^2+\half(\lambda_{13}\eta^2+\lambda_{23}\varphi^2)-\mu_3^2\,,\\
m_{G_0}^2 &= m_{G^{\pm}}^2 = \lambda_3 h^2+\half(\lambda_{13}\eta^2+\lambda_{23}\varphi^2)-\mu_3^2\,,\\
m^2_{W^{\pm}} &=\frac{1}{4} g^2 h^2\,,\\
m^2_t &= \frac{1}{2} y_t^2 h^2\,.
\end{align}

The neutral gauge bosons $(W_3,~B)$, mix to yield the mass matrix
\beq
\mathcal{M}^2_{Z\gamma} = \frac{1}{4}\begin{bmatrix}
g^2h^2 & -g g' h^2\\
-g g' h^2 & g'^2(h^2+4\varphi^2) 
\end{bmatrix}
\eeq
Diagonalization gives the field-dependent masses for the $Z$ boson and the photon, $\gamma$.

The bosonic thermal squared masses $\Pi_b$, at leading order, are given by
\begin{align}
    \Pi_{\eta} &= C_{\eta}^0 T^2\,,\\
    \Pi_{\varphi} &= \Pi_{\xi} = C_{\varphi}^0 T^2\,,\\
    \Pi_h &= \Pi_{G_0} = \Pi_{G^{\pm}} = C_h^0 T^2 \,,\\
    \Pi^L_{W^i} &= \frac{11}{6} g^2T^2\,,~(i=1,2,3)\,,\\
    \Pi^L_{B} &= \frac{19}{6} g'^2T^2\,,
\end{align}
where the coefficients $C^0_\eta$, $C^0_\varphi$, and $C^0_h$ are given in Eqs.\,\eqref{eq: C0_eta},\,\eqref{eq: C0_varphi},\,\eqref{eq: C0_h} respectively.

\bibliographystyle{JHEP}
\bibliography{arxiv_v1.bbl}

\end{document}